%% file: tmag_sot_cam_5T.tex
\documentclass[journal]{IEEEtran}

\usepackage{cite}
\usepackage{amsmath,amssymb,amsfonts}
\usepackage{algorithmic}
\usepackage{graphicx}
\usepackage{textcomp}
\usepackage{xcolor}
\def\BibTeX{{\rm B\kern-.05em{\sc i\kern-.025em b}\kern-.08em
    T\kern-.1667em\lower.7ex\hbox{E}\kern-.125emX}}

\begin{document}

\title{A 5T-2MTJ STT-assisted Spin Orbit Torque based Ternary Content Addressable Memory for Hardware Accelerators}

\author{Siri~Narla,\thanks{This work was supported in part by CoCoSys, one of seven centers in JUMP 2.0, a Semiconductor Research Corporation (SRC) program sponsored by DARPA.}\thanks{Siri Narla, Piyush Kumar, and Azad Naeemi are with the School of Electrical and Computer Engineering, Georgia Institute of Technology, Atlanta, GA 30332 USA, email: snarla6@gatech.edu.}~Piyush~Kumar,~\IEEEmembership{Student~Member,~IEEE},~and~Azad~Naeemi,~\IEEEmembership{Senior~Member,~IEEE}}

\maketitle

\input{tex/abstract}
\input{tex/intro}
\input{tex/cell}
\input{tex/write}
\input{tex/search}

\input{tex/results}

\input{tex/conclusion}
\input{tex/ack}

\bibliographystyle{unsrt}
\bibliography{sample-base}

\end{document}

%% file: tex/abstract.tex
\begin{abstract}

In this work, we present a novel non-volatile spin transfer torque (STT) assisted spin-orbit torque (SOT) based ternary content addressable memory (TCAM) with 5 transistors and 2 magnetic tunnel junctions (MTJs). We perform a comprehensive study of the proposed design from the device-level to application-level. At the device-level, various write characteristics such as write error rate, time, and current have been obtained using micromagnetic simulations. The array-level search and write performance have been evaluated based on SPICE circuit simulations with layout extracted parasitics for bitcells while also accounting for the impact of interconnect parasitics at the 7nm technology node. A search error rate of 3.9x10\textsuperscript{-11} is projected for exact search while accounting for various sources of variation in the design. In addition, the resolution of the search operation is quantified under various scenarios to understand the achievable quality of the approximate search operations. Application-level performance and accuracy of the proposed design have been evaluated and benchmarked against other state-of-the-art CAM designs in the context of a CAM-based recommendation system.

\end{abstract}

%% file: tex/intro.tex
\section{Introduction}
With the unprecedented growth of retrieval based applications in a diverse range of areas from recommendation systems for e-commerce \cite{sasrec} and video or image retrieval applications \cite{Noh_2017_ICCV} to search engines \cite{faiss, disk_ann} and genomics \cite{pearson2013introduction}, performing accurate and fast similarity search has become increasingly important since it is a vital function for all these use cases. Graph-based similarity search algorithms \cite{wang2021comprehensive} have been extensively studied and have proven to be effective for low-dimensional data. However, these algorithms become inefficient as the dimensionality of the data increases \cite{fu2018fast}. In high-dimensional applications, such as image retrieval, hash encoding \cite{luo2023survey, wang2014hashing} is often employed to improve performance. Despite these advancements, both graph-based and hashing-based algorithms face challenges with scalability, as their execution time increases significantly with the number of samples. In modern applications, dataset sizes can be large in both dimensions (D) and number of samples (n), because of which similarity search becomes increasingly challenging \cite{tong2008lessons}. Thus, hardware solutions like content addressable memories (CAMs), that can perform parallel in-memory search over an entire database, are being studied with great interest \cite{narla2022design, narla2022modeling,  yin2018ultra}. The speed-up in search they offer and their linear storage requirements have made CAMs suitable for applications such as memory augmented neural networks \cite{fefet_mann}, hyper-dimensional computing \cite{HD_intro}, recommendation systems \cite{iMARS}, and dataset searches.

Many designs and devices have been explored to implement CAMs. Traditional CMOS implementations of TCAM (ternary content addressable memories) using SRAMs \cite{pagiamtzis2006content} require 16 transistors which limits the memory capacity. SRAMs also suffer from leakage which can become a major limiting factor for large data-sets and for applications with large idle times \cite{narla2022modeling}. To address these challenges, CAM cells based on emerging non-volatile memory (NVM) devices such as resistive memories (RRAMs) \cite{li20131mb}, ferroelectric field effect transistors (FeFETs) \cite{yin2018ultra}, and spin-orbit torque (SOT) devices \cite{narla2022design}, have been studied. Each of these technology options come with their own set of advantages and limitations. RRAM-based TCAMs \cite{2t2r_tcam} need very sensitive sense amplifiers due to low sensing margin. PCMs suffer from resistance drift issues \cite{rdrift} and require large write voltages. FeFETs are voltage-controlled and can be quite energy efficient and compact. However, they generally suffer from poor endurance \cite{yurchuk2016charge}, have a finite read after write \cite{finite_read1,finite_read2}, and generally require large write voltages (4V) \cite{yin2018ultra}. SOT devices theoretically have unlimited endurance, good retention times \cite{narla2022design}, and compatibility with CMOS back-end-of-line (BEOL) \cite{garello2019manufacturable}. Large scale wafer-level implementation of SOT devices have already been demonstrated \cite{garello2019manufacturable,song2022high}. Although, they need larger write energy than FeFET and CMOS-based CAMs due to their current based write scheme \cite{narla2022design}, most CAM-based applications are more search-intensive with a limited number of writes. 

\par

For SOT devices, the magnetic orientation of the ferromagnetic free layers is determined by their anisotropy direction and they can either have in-plane magnetic anisotropy (IMA) or perpendicular magnetic anisotropy \cite{mangin2006current}. The IMA based devices rely on the aspect ratio of the ferromagnet to maintain sufficient thermal stability and retention time \cite{chun2012scaling}. It results in IMA devices being hard to scale to advanced technology nodes as the thermal stability becomes very sensitive to lithography variations. On the other hand, PMA devices rely on interface anisotropy \cite{ikeda2010perpendicular} for thermal stability and are easier to scale to advanced nodes.
Despite these advantages, one major concern with PMA based SOT devices is that they rely on external magnetic field to achieve the required symmetry breaking for deterministic write operation \cite{garello2019manufacturable}.
While there has been research to mitigate this issue by using magnetic hard mask \cite{garello2019manufacturable} or exchange bias \cite{fukami2016magnetization}, these solutions may have further challenges related to scaling to advanced nodes and may sacrifice magnetic immunity \cite{lin2021challenges}.
To solve this issue, we use STT-assisted SOT switching which eliminates the need for a magnetic field for the write operation. Also, compared to the previous SOT-CAM \cite{narla2022design, narla2024cross} implementations, our current design does not require any changes to the bit-cell to enable ternary storage capabilities.

 \par
The rest of the paper is organized as follows. After this introduction, we discuss the cell design in section II. Section III discusses the write operation and Section IV discusses the exact and approximate search operations, search error rate, and resolution. In Section V we evaluate and benchmark our design against other CAMs at the 7nm technology node using resolution, energy, and delay as metrics. In addition, we benchmark the design at the application-level using a CAM-based recommendation system. Finally, Section VI concludes the paper.

%% file: tex/cell.tex
\section{Cell design and operation}
\begin{figure}
	\centering
	\includegraphics[width=\linewidth]{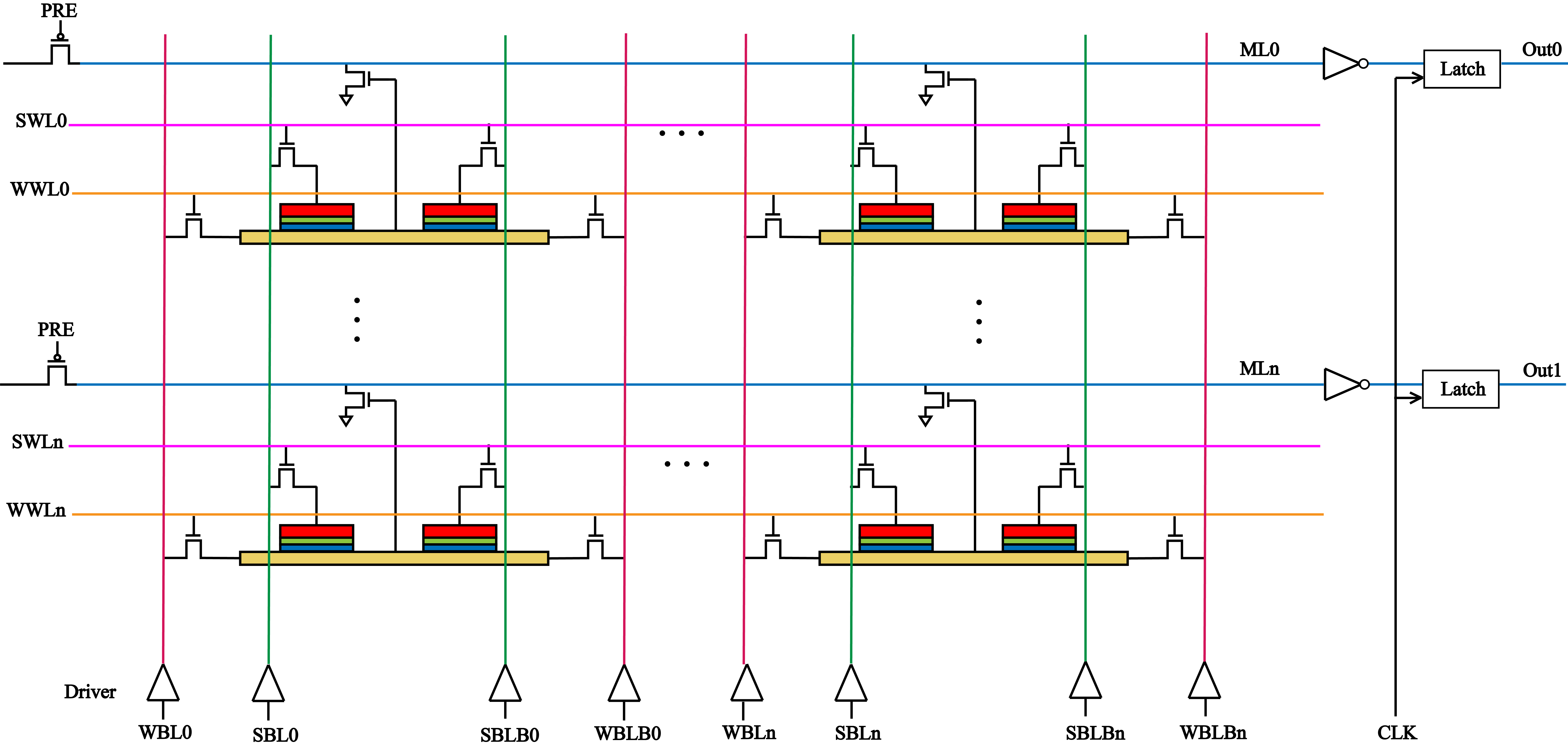}
	\caption{SOT-TCAM array with search drivers, precharge transistors, and sensing inverter amplifiers.}
	\label{fig_tcam_array}
\end{figure}

The proposed TCAM cell consists of 5 transistors and 2 MTJs as shown in Fig. \ref{fig_tcam_cell_schematic}. It can store and search for `0', `1', and `don’t care' (`X') bits. Fig. \ref{fig_tcam_array} shows how these cells can form a TCAM array.

\subsection{Write operation}
The write operation is based on STT-assisted SOT switching of perpendicular magnets \cite{jxcdc_sot_stt_vcma}. In this scheme, first an SOT current is applied for a small duration of time (~1 ns) which causes the magnetizations of the free layers of the two MTJs to move towards the in-plane meta-stable state. After that, an STT current of appropriate polarity is applied for deterministic switching. Table \ref{tab:write_operation} shows the various write voltage values along with the MTJ states for writing `0', `1', and `X' bits. The detailed discussion of the simulation methodology used to derive these values is described in Section III.

\subsection{Search operation}
\begin{figure}
	\centering
	\includegraphics[width=\linewidth]{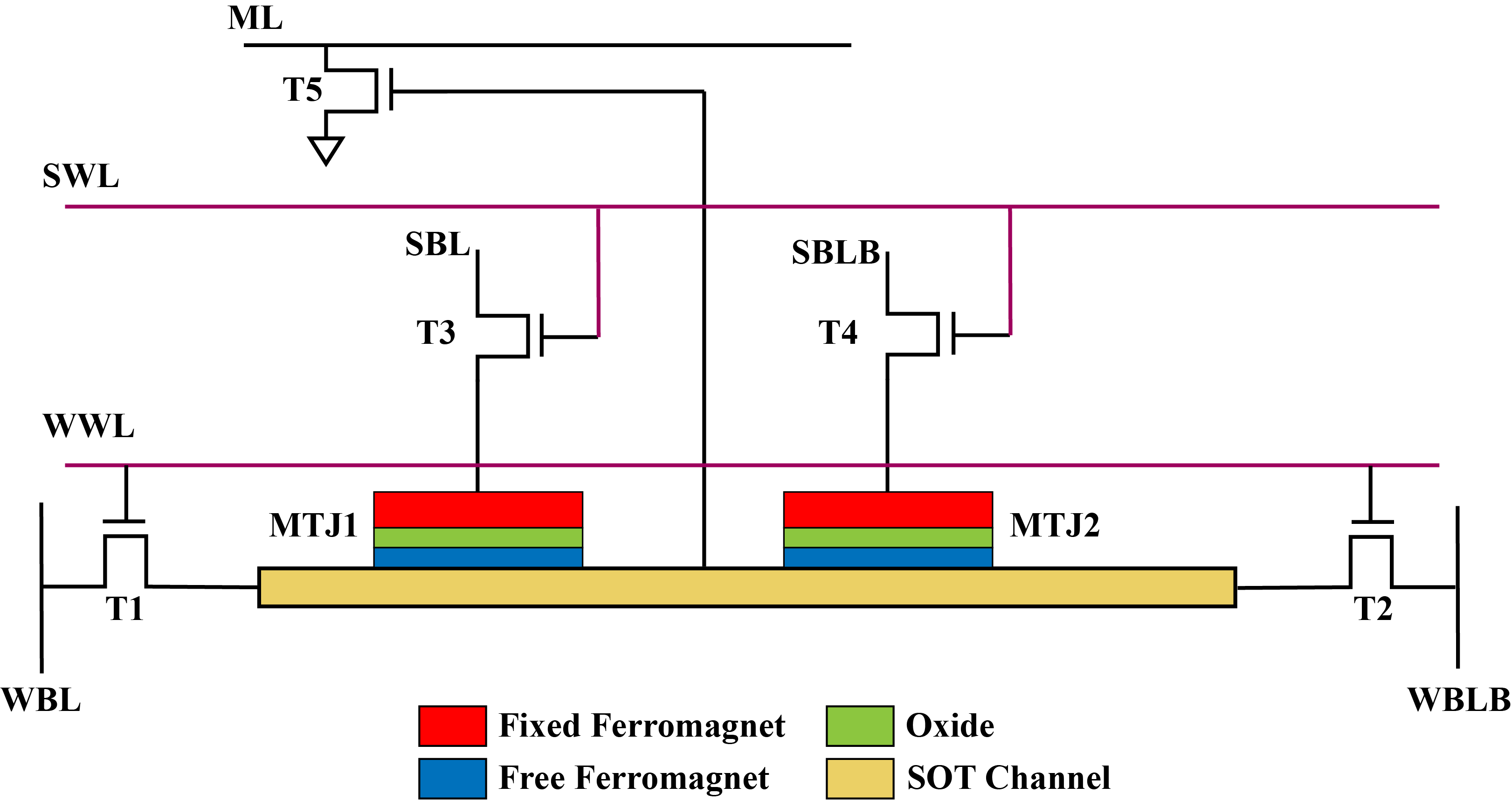}
	\caption{5T-2MTJ SOT-TCAM cell schematic. T1 and T2 are SOT write transistors. T3 and T4 are STT/search transistors. T5 is the discharge transistor for the ML. WWL: write wordline, SWL: search wordline, ML: matchline, WBL: write bitline, SBL: search bitline.}
	\label{fig_tcam_cell_schematic}
\end{figure}

The search operation is based on discharging the matchline (ML) if there is a mismatch between the stored data and the search data. ML is precharged before every search operation. During the search operation, WWL is kept off, while SWL is switched on and the complementary search data is applied via SBL and SBLB. The voltage ($\mathrm{V_{sot}}$) at the gate of the discharge transistor (T5 in Fig. \ref{fig_tcam_cell_schematic}) is determined by the voltage division of the search voltage ($\mathrm{V_{s}}$) between the two MTJs. In the case of a mismatch, $\mathrm{V_{sot}}$ is designed to be larger than the threshold voltage of the discharge transistor which results in ML getting discharged through T5. If the stored bit is `don't care' ('X'), or there is a match, $\mathrm{V_{sot}}$ is designed to be adequately lower than the threshold voltage of T5 to prevent ML discharge. If the search bit is `X', SBL and SBLB are grounded and T5 remains off regardless of the stored data. Table \ref{tab_read} summarizes the search operation for all possible stored bit and search bit combinations. Every ML whose output at a certain point in time is stored in an output latch. By adjusting this time period, the latched data can represent the rows within a certain hamming distance from the search data  as will be discussed in Section IV.


\begin{table}[]
    \caption{Various WL and BL voltage values used during the write operation. HRS - high resistance state, LRS - low resistance state.}
    \label{tab:write_operation}
    \centering
    \begin{tabular}{r|cccc}
        \hline
        \hline
        Write data  & MTJ1/MTJ2 & WWL/SWL & WBL/WBLB & SBL/SBLB \\
        \hline
        `1' - SOT & HRS/LRS & 1.56V/0V & 1.56V/0V & 0V \\
        - STT & ~ & 1.56V & 0.42V & 1.56V/0V \\
        \hline
        `0' - SOT & LRS/HRS & 1.56V/0V & 1.56V/0V & 0V \\
        - STT & ~ & 1.56V & 0.42V & 0V/1.56V \\
        \hline
        `X' - SOT & HRS/HRS & 1.56V/0V & 1.56V/0V & 0V \\
        - STT & ~ & 1.56V & 0.42V & 1.56V \\
        \hline
    \end{tabular}
\end{table}

\begin{table}
    \caption{TCAM search operation. $\mathrm{k=R_p/(R_p+R_{ap})}$. $\mathrm{R_p}$ and $\mathrm{R_{ap}}$ are the MTJ resistances in parallel and antiparallel states. For storing '1', MTJ1=$\mathrm{R_p}$ and MTJ2=$\mathrm{R_{ap}}$. For storing '0', MTJ1=$\mathrm{R_{ap}}$ and MTJ2=$\mathrm{R_p}$. For storing '1', MTJ1=$\mathrm{R_{ap}}$ and MTJ2=$\mathrm{R_{ap}}$. }
	\label{tab_read}
	\centering
	\begin{tabular}{c|ccccc}
		\hline
		\hline
		{ } & \textbf{Stored data} & \textbf{SBL} & \textbf{SBLB} & $\mathbf{V_{sot}}$ & \textbf{ML} \\
		\hline
		\textbf{Searching `1'} & 1 & $\mathrm{V_s}$ & 0 & k$\mathrm{V_s}$ & High \\
		{} & 0 & $\mathrm{V_s}$ & 0 & (1-k)$\mathrm{V_s}$ & Low \\
		{} & X & $\mathrm{V_s}$ & 0 & $\mathrm{V_s}$/2 & High \\
		\hline
		\textbf{Searching `0'} & 1 & 0 & $\mathrm{V_s}$ & (1-k)$\mathrm{V_s}$ & Low \\
		{} & 0 & 0 & $\mathrm{V_s}$ & k$\mathrm{V_s}$ & High \\
		{} & X & 0 & $\mathrm{V_s}$ & $\mathrm{V_s}$/2 & High \\
		\hline
		\textbf{Searching `X'} & 1 & 0 & 0 & 0 & High \\
		{} & 0 & 0 & 0 & 0 & High \\
		{} & X & 0 & 0 & 0 & High \\
		\hline
	\end{tabular}
\end{table}

\subsection{Layout and parasitic extraction} 
\begin{figure}
    \centering
    \includegraphics[angle=90, width=3.3in]{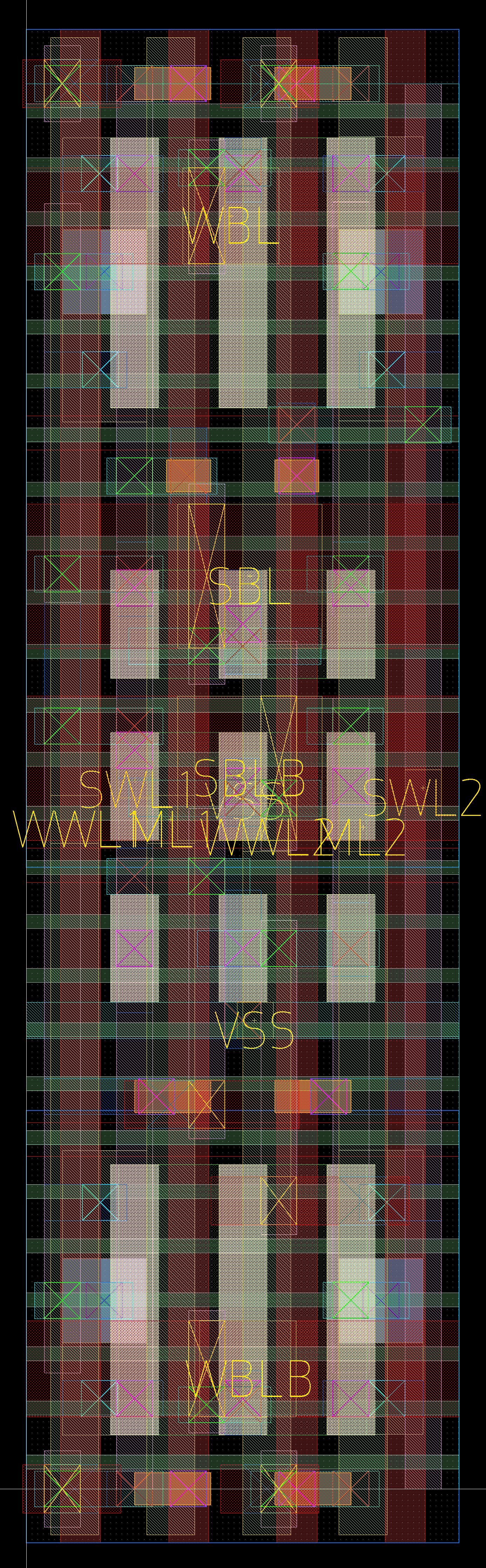}
    \caption{Layout for the SOT-TCAM cell with 2 bit-cells sharing the BL contacts. Effective area of a single bit-cell is 0.108um x 0.702um.}
    \label{fig:cam_layout}
\end{figure}

To accurately estimate the impact of various sources of parasitics in the circuit, we design the layout of the cell at the 7nm node using the ASAP7 PDK \cite{clark2016asap7}, where we assume an optimistic fin height of 52 nm. Fig. \ref{fig:cam_layout} shows the layout of 2 adjacent bit-cells sharing the bitline (WBL/WBLB/SBL/SBLB) contacts among them. The SOT layer and MTJ are placed between the M1 and M2 metal layers. The proposed bit-cell uses metal levels M1-M5. The write wordline (WWL) and matchline (ML) are routed on M5 while the search wordline (SWL) is routed on M3. Write bitlines (WBL/WBLB) and search bitlines (SBL/SBLB) are routed on M4. Since the performance and accuracy of the cell depend significantly on the interconnect resistance \cite{jxcdc_mem_intc}, we use wider wires for the bitlines. For WBL/WBLB, we use a wire width of 48 nm which is 2x the minimum wire width of M4, while SBL/SBLB use 72 nm wide wires (3x minimum width of M4). The effective bit-cell area is 0.076 um2.

To extract the cell parasitics, the BEOL resistance values from \cite{huang2023comprehensive} are used to generate the nxtgrd database containing the resistance and capacitance information for various layers. This nxtgrd database is then used in Synopsys’ StarRC for extracting the parasitic resistance and capacitance from the layout.
These parasitics are then used in SPICE simulations to measure ML discharge delays for various array sizes, Hamming distance (HDist) values and row positions. For SOT-CAMs we use an MTJ low resistance state of 25 k$\Omega$, a tunnel magnetoresistance ratio (TMR) of 1.8, and half bias values based on experimental results from \cite{yuasa2004giant}. The TMR degradation due to bias voltage is also incorporated as shown in \cite{narla2022design}. In the previous SOT-CAM designs \cite{narla2022design, narla2024cross} a much larger LRS resistance (1 M$\Omega$) was used to reduce the search energy. However, in this design, we have to use LRS resistance = 25 k$\Omega$, since for larger LRS values the current through the MTJ is too small to write to the device. Resistance area product of an MTJ depends on its oxide thickness, hence by changing the oxide thickness value we can change the MTJ resistance.

%% file: tex/write.tex
\section{Write simulations}

\begin{figure}
    \centering
    \includegraphics[width=\linewidth]{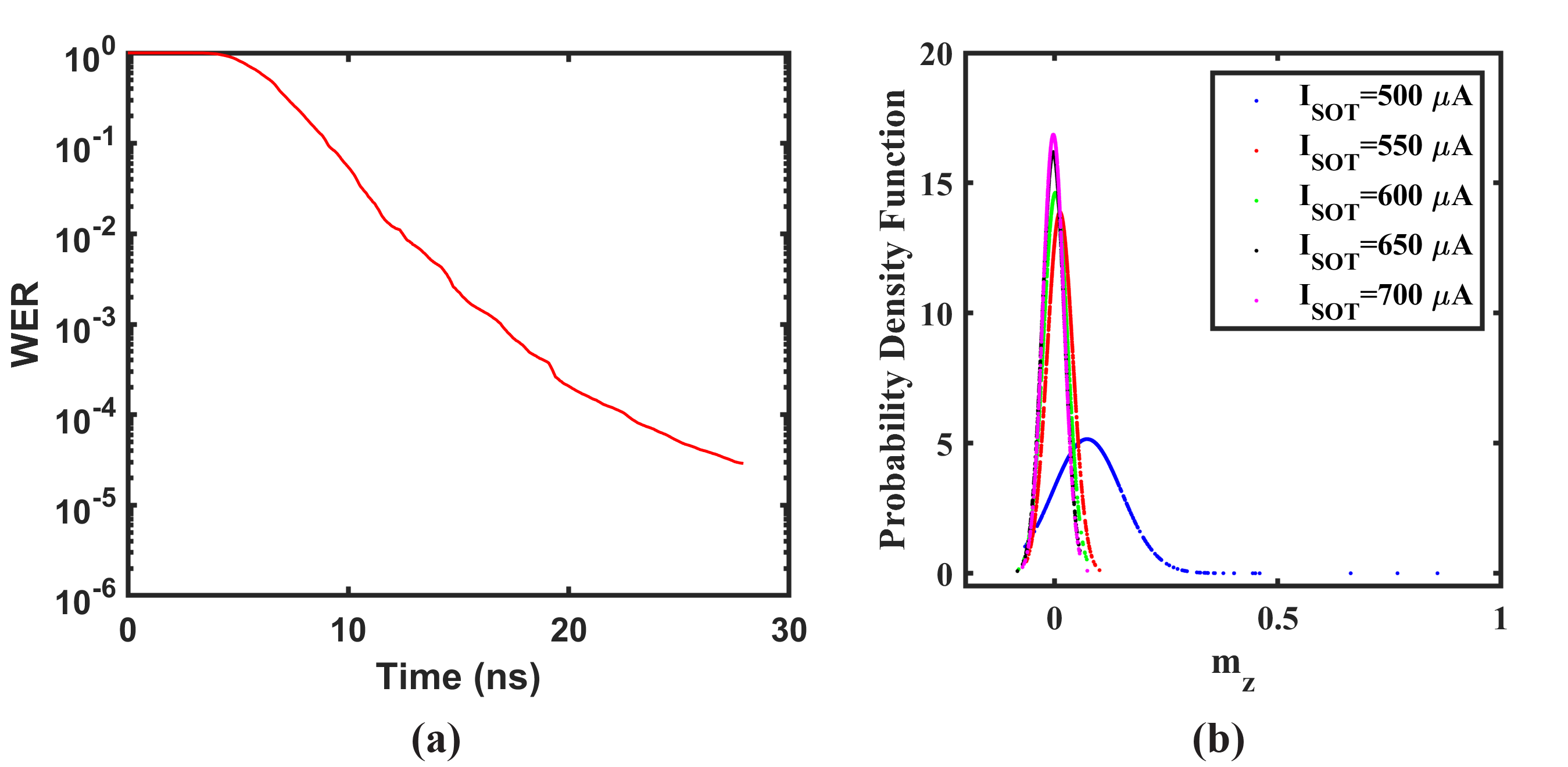}
    \caption{(a) Write error rate plotted against the write delay for STT-assisted SOT switching. SOT current is applied for the initial 1 ns. (b) Distribution of mz at the end of 1 ns SOT pulse for varying values of SOT spin current.}
    \label{fig:wer}
\end{figure}

To obtain the minimum required duration of the STT current required, we perform OOMMF \cite{oommf2} based micromagnetic simulations augmented with rare event enhancement \cite{roy2016write}. In the past, we have validated the results of these kinds of simulations with experimental data \cite{kumar2022benchmarking}. Fig. \ref{fig:wer}(a) shows the write error rate (WER) obtained for an STT spin current of 6 uA. To obtain WER of 1e-5, we use a write time of 30 ns. Increasing the STT current can reduce the write time; however, it requires lowering the MTJ resistance which negatively impacts the search performance and accuracy as will be explained in Section IV. Thus, we choose a small enough value for STT current which can reliably achieve low WER in the presence of thermal noise. While the write speed for SOT+STT switching is slower compared to the field-assisted SOT case ($\sim$1-2 ns), the former has better magnetic immunity than the later. The spin current generated from SOT is 700 uA and the SOT layer considered here is 3.5 nm thick $\beta-$W \cite{imec_beta_w}. The resistivity and damping-like spin torque efficiency values used for $\beta-$W are 160 $\mu \Omega-$cm and 0.33, respectively \cite{imec_beta_w}. The magnitude of SOT current does not have any impact on WER as long as it is sufficient to drive the magnetization towards the in-plane meta-stable state. Fig. \ref{fig:wer}(b) shows the distribution of the z component of magnetization ($m_{z}$) at the end of 1 ns SOT pulse for varying values of SOT spin current. For SOT spin current below 600 $\mu$A the $m_{z}$ distribution starts to widen and too much deviation from $m_{z}$=0 can lead to switching failures. For the free layer ferromagnet, we consider 1.3 nm thick CoFeB with room temperature saturation magnetization of 1.2 MA/m and interface anisotropy of 1.15 mJ/m2 which gives a room temperature thermal stability of 65. The MTJ diameter is 60 nm. In addition, we assume that the STT efficiency is 0.6 for anti-parallel (AP) to parallel (P) switching and 0.3 for P to AP switching \cite{jaiswal2016comprehensive}.


The required electric write currents are 210 uA and 10/20 uA for SOT and STT phases, respectively. We choose the write voltages by adding 10\% margin over the required currents.
To achieve a write current of 20 uA with 1.56V supply, the MTJ resistance in the parallel state has to be 25 k$\Omega$ or smaller. Since the search function can be more energy efficient if the MTJ resistance is larger. Hence, we chose the largest possible value.
Table \ref{tab:write_operation} describes the various voltages used during the write operation. The direction of the SOT current is fixed and is applied by driving WBL to 1.56V and WBLB to ground while keeping the WWL on. During the STT phase, both WBL/WBLB are driven to 0.42V while applying either 1.56V or 0V on SBL/SBLB to write the appropriate data. During the STT phase, both SWL and WWL remain enabled. Based on full array simulation, we obtain write energies of 1.74 pJ per bit for writing binary data (0/1) and 3.05 pJ per bit for writing `X' for an array size of 64x64.


%% file: tex/search.tex
\section{Search simulations}

\par
The discharge rate of the ML increases with Hamming Distance which is equal to the number of mismatching bits between the two vectors. For a row whose stored data perfectly matches the search data, the voltage on the ML remains high since all the discharge transistors in the row are off and the only current discharging the ML is the subthreshold leakage currents from these transistors. Thus, finding a fully matching vector requires identifying the ML with discharge rate slower than the worst-case mismatch, i.e. a row with single bit mismatch.
On the other hand, finding the vector closest to a search query requires identifying the ML with the slowest discharge rate.
Hence, the discharge rate becomes an important factor for approximate search (or nearest neighbor search). 
\par
Fixed-radius near neighbor search is the search of all items with an HDist smaller than a given value (HDist limit). 
It can be implemented using delay thresholding with a latch connected to the output of the inverters connected to the match lines and by controlling the timing of the clock (Clk) falling edge at the latches.
All rows whose ML discharge remain undetected by their inverters before the Clk falling edge arrival are considered to be neighbors within the specified radius.

\par

While performing search, each search line in the array has a driver that drives the column to the associated search voltage. Due to the low LRS value used, the effective resistance from all the rows between the SBL/SBLB drivers becomes comparable to the driver resistance. Thus, the voltage drops across the search drivers are more pronounced which reduces the search voltage window available for voltage division between the MTJs within a cell. A large number of fins on the driver transistors can mitigate this effort by lowering the driver resistance. As the number of rows in an array increases, the search voltage window further shrinks and makes the design more vulnerable to non-idealities. Hence, to ensure good search capabilities, we need to limit the number of rows to 64.

\subsection{Exact Match Search Error Rate}
\begin{figure}[t]
	\centering
	\includegraphics[width=0.8\linewidth]{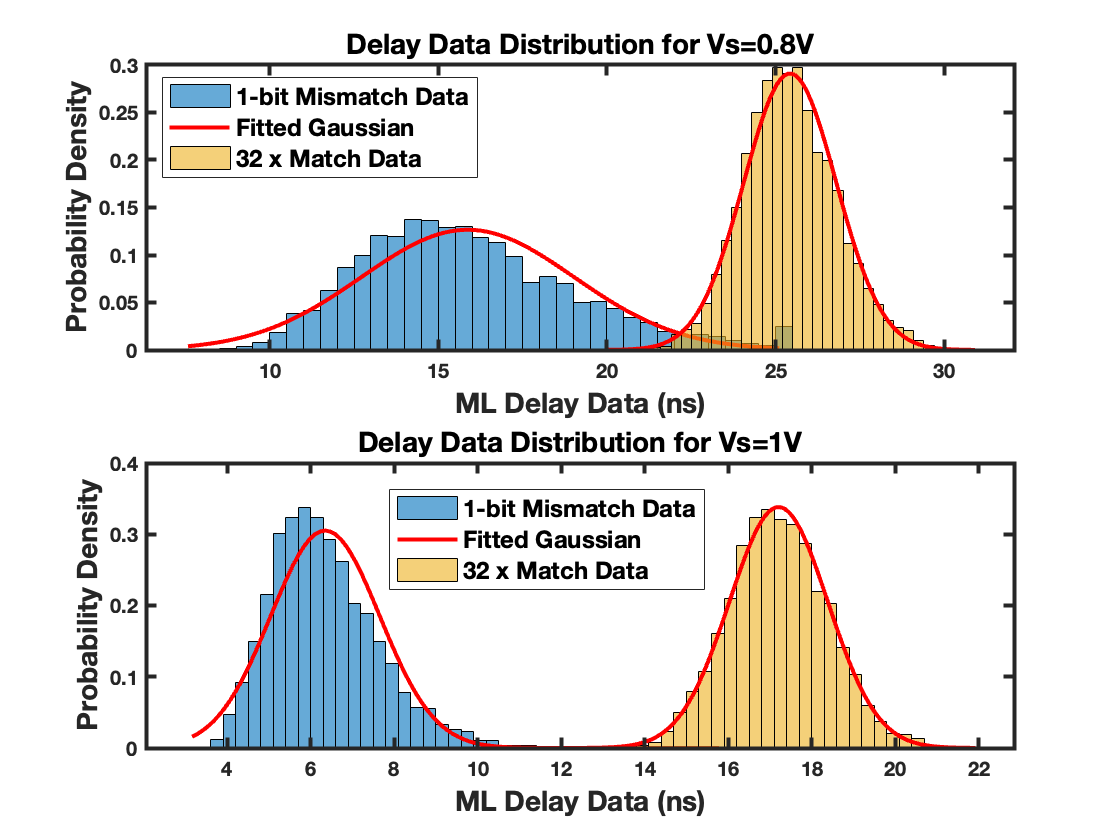}
	\caption{ML delay data distribution for 1000 Monte Carlo simulations, considering Vt and MTJ resistance variation, for single bit mismatch and 32 `X' bit match for a 64x128 SOT-5T CAM array.}
	\label{fig_ser}
\end{figure}

To evaluate the robustness of our design during the search operation, we first calculate the search error rate (SER) for detecting an exact match.
A search error occurs when the off-state leakages of the transistors, due to magnetic and CMOS device variability, discharge the ML of a `match row' at a faster rate than that of a `mismatch row', thus sensing a match as a mismatch. Stored `don't care' bits in a row with full match contribute to more leakage than stored binary bits because `don't care' bits have a larger $\mathrm{V_{sot}}$ value ($\sim$$\mathrm{V_{sot}}$/2). Hence, we focus on the case of sensing a match on a row storing 32 `X' bits in a 128-bit vector to better see the impact of the device parameters on the SER. We use 3$\sigma$ MTJ resistance variation of 15$\%$ \cite{everspin} and a 3$\sigma$ transistor threshold voltage variation of 42 mV \cite{giles2015high}. We assume Gaussian distribution and run 1000 Monte Carlo simulations for various row positions in the array. 
We obtain the ML delay distributions corresponding to full match with 32 stored `X' bits and worst case mismatch with one mismatching bit.
The resulting delay data is fitted with a Gaussian distribution to calculate the SER (overlapping area between the two curves) as shown in Fig. \ref{fig_ser}.
For $\mathrm{V_s}$ = 0.8 V and 1 V, we achieve the SER values of 1.53\% and 3.9x10\textsuperscript{-9}\%, respectively. Increasing the search voltage increases the difference between the $\mathrm{V_{sot}}$ for mismatch ((1-k)$\mathrm{V_s}$) and `X' bit match ($\mathrm{V_s}/2$); hence, improving the design's variation tolerance at the expense of increased power consumption.


\subsection{Minimum Detectable Distance (Resolution)}
\begin{figure}[t]
	\centering
	\includegraphics[width=0.8\linewidth]{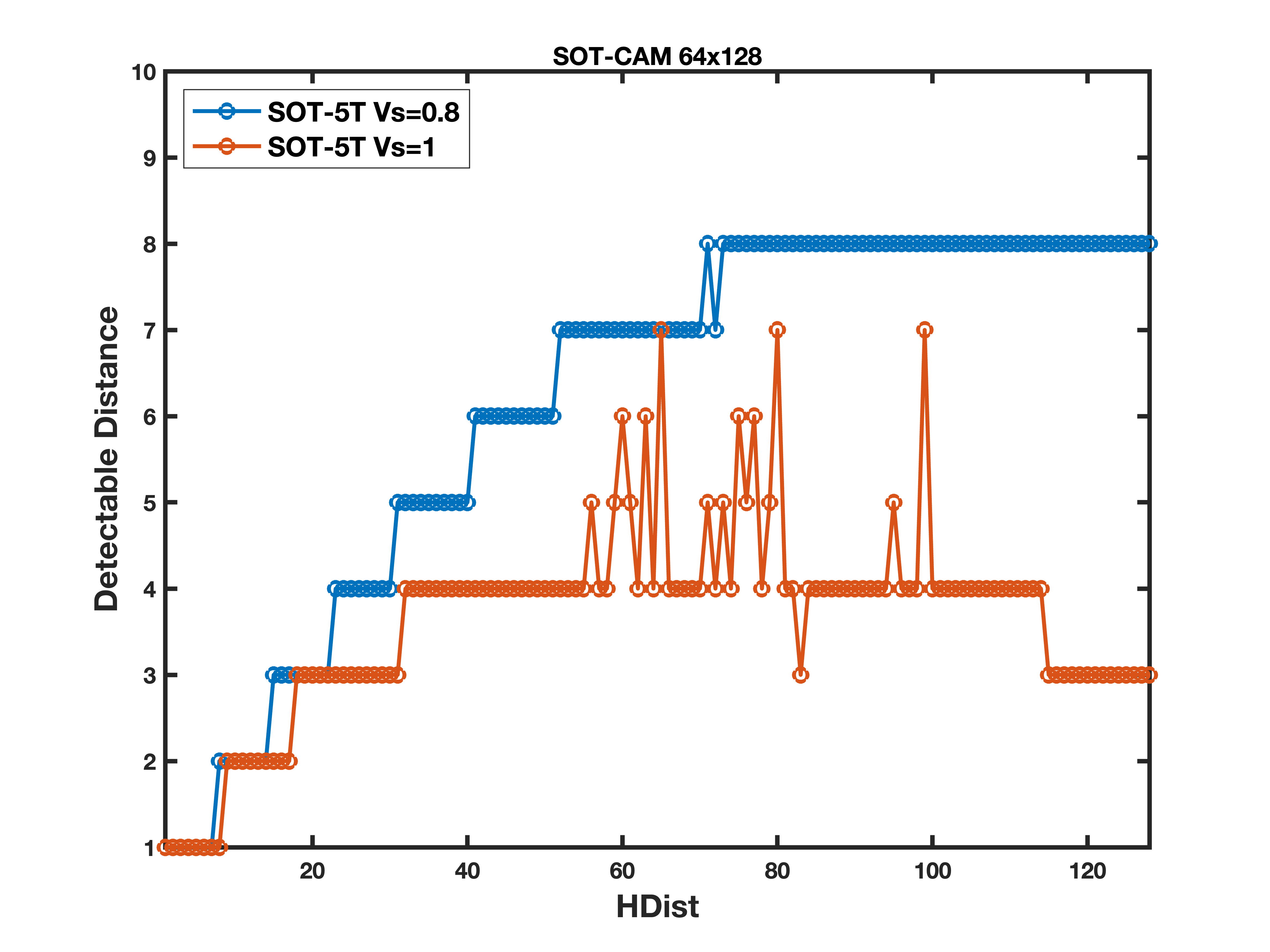}
	\caption{Minimum detectable distance for 5T SOT-CAM with Vs=0.8 V and 1 V.}
	\label{fig_mdd_5t}
\end{figure}

Search resolution is the ability to clearly distinguish between the rows with similar Hamming distance values. This can be measured in terms of the minimum detectable distance (MDD). If the discharge delay distribution corresponding to HDist=n have no overlap with the discharge delay distribution corresponding to HDist=(n $\pm$ $\delta$), then MDD = $\delta$. MDD is plotted in Fig. \ref{fig_mdd_5t} for every HDist for $\mathrm{V_s}$ = 0.8 V and 1 V. 
\par
Due to the low LRS resistance value, the SOT-5T case has a large current flowing through SBL/SBLB. This contributes to a substantial IR drop over the SBL/SBLB because of which the $\mathrm{V_{sot}}$ mismatch for the row furthest from the driver is lower than the $\mathrm{V_{sot}}$ mismatch for the row closest to the driver. Widening the SBL/SBLB wires helps in reducing the IR drop.

\par
Increasing the search voltage to 1V increases $\mathrm{V_{sot}}$ for mismatch. A larger mismatch $\mathrm{V_{sot}}$ value reduces the overall delay due to larger discharge current. Hence, the difference in delays for consecutive Hamming distances decreases. However, since the ML delay is inversely proportional to the discharge current, the impact of $\mathrm{V_{sot}}$ mismatch on delay variation at various row positions is reduced for Vs=1 V compared to Vs=0.8 V. Thus, the resolution improves with an increase in the search voltage as seen in Fig. \ref{fig_mdd_5t} due to lower variation in the delay values for a fixed HDist.
\subsection{Fixed-radius Search}
\par
Table \ref{tab_def} defines the metrics used to evaluate the accuracy for a fixed-radius search operation. Table \ref{tab_nns_5t} shows the results for a fixed radius search operation for a Hamming distance threshold of 20 over a randomly generated dataset with 10000 128-bit vectors stored over multiple CAM arrays of size 64x128. For the ideal case in Table \ref{tab_nns_5t}, we ignore interconnect parasitics. 
We see a drop in precision for the realistic case since the rows (with HDist $>$ 20) further away from the search drivers have ML delays larger than the threshold delay used to capture all the rows with Hamming distances $\leq$ 20. Precision is closely related to the resolution of the CAM design. Since Vs=1 V has higher resolution, it shows better precision. We use this fixed radius search to implement a CAM-based recommendation system to analyze results at the application-level in Section V.

\begin{table}[]
	\caption{Precision, Recall and F-score definition}
	\label{tab_def}
	\begin{tabular}{l|l}
		\hline
        \hline
		Precision & Percentage of relevant items among the retrieved items \\ 
		Recall    & Percentage of relevant items that were retrieved       \\ 
		F-score   & Harmonic mean of precision and recall                  \\ \hline
	\end{tabular}
\end{table}
 
\begin{table}[t]
    \caption{fixed radius search with HDist threshold=20 for 5T SOT-CAM with Vs=0.8V and 1V.}
    \label{tab_nns_5t}
    \centering
    \begin{tabular}{l|cccc}
    \hline
        \hline
        & Ideal $V_s = 0.8$ & $V_s = 0.8$ & Ideal $V_s = 1$ & $V_s = 1$ \\
        \hline
        Recall (\%) & 100 & 100 & 100 & 100 \\
        Precision (\%) & 100 & 93.52 & 100 & 97.03 \\
        F-score & 1 & 0.97 & 1 & 0.985 \\
        \hline
    \end{tabular}
\end{table}

%% file: tex/results.tex
\begin{table*}   
    \caption{fixed radius search with HDist threshold=20 for 5T, 3T SOT-, FeFET and SRAM-based CAM.}
    \label{tab_nns_bench}
    \centering
    \begin{tabular}{l|cccccc}
    \hline
    \hline
        & \textbf{SOT-5T} (0.8V) & \textbf{SOT-5T} (1V) & \textbf{SOT-3T} (0.8V) & \textbf{SOT-3T} (1V) & \textbf{SRAM} & \textbf{FeFET} \\
        \hline
        Recall (\%) & 100 & 100 & 100 & 100 & 100 & 100 \\
        Precision (\%) & 93.52 & 97.03 & 96.88 & 94.9 & 100 & 93.13 \\
        F-score & 0.97 & 0.985 & 0.984 & 0.974 & 1 & 0.96 \\
        \hline
    \end{tabular}
\end{table*}

\begin{table*}[]
    \caption{Benchmarking results for a CAM-based Recommendation System.}
    \label{tab_sasrec_bench}
    \centering
    \begin{tabular}{l|cccccc}
    \hline
    \hline
        & \textbf{SOT-5T} (0.8V) & \textbf{SOT-5T} (1V) & \textbf{SOT-3T} (0.8V) & \textbf{SOT-3T} (1V) & \textbf{SRAM} & \textbf{FeFET} \\
        \hline
        HR@10 & 0.32 & 0.32 & 0.32 & 0.32 & 0.32 & 0.32 \\
        Mean Pool Size & 230.75 & 188.45 & 225.35 & 219.1 & 219.5 & 484.7 \\
        DPR reduction & 4.33x & 4.88x & 4.1x & 4.58x & 4.56x & 2.06x \\
        \hline
    \end{tabular}
\end{table*}

\begin{table*}[]
    \caption{Search energy and delay results for fixed radius search for Hamming distance of 10 using a 64x128 CAM array.}
    \label{tab_edp_exact}
    \centering
    \begin{tabular}{l|cccccc}
    \hline
    \hline
        & \textbf{SOT-5T} (0.8V) & \textbf{SOT-5T} (1V) & \textbf{SOT-3T} (0.8V) & \textbf{SOT-3T} (1V) & \textbf{SRAM} & \textbf{FeFET} \\
        \hline
        Delay (ns) & 1.6 & 0.7 & 1.6 & 1.27 & 0.72 & 1.17 \\
        Energy (pJ) & 56.4 & 43.4 & 3.24 & 6.17 & 0.716 & 1.22 \\
        \hline
    \end{tabular}
\end{table*}

\begin{table*}[]
    \caption{Search energy and delay results for exact search using a 64x128 CAM array.}
    \label{tab_edp_approx}
    \centering
    \begin{tabular}{l|cccccc}
    \hline
    \hline
        & \textbf{SOT-5T} (0.8V) & \textbf{SOT-5T} (1V) & \textbf{SOT-3T} (0.8V) & \textbf{SOT-3T} (1V) & \textbf{SRAM} & \textbf{FeFET} \\
        \hline
        Delay (ns) & 12 & 5.78 & 4.7 & 2.83 & 4.8 & 7.6 \\
        Energy (pJ) & 433 & 366 & 9.17 & 13.1 & 1.15 & 1.89 \\
        Cell Area (um sq) & 0.076 & 0.076 & 0.058 & 0.058 & 0.109 & 0.044 \\
        \hline
    \end{tabular}
\end{table*}

\begin{figure}
	\centering
	\includegraphics[width=0.8\linewidth]{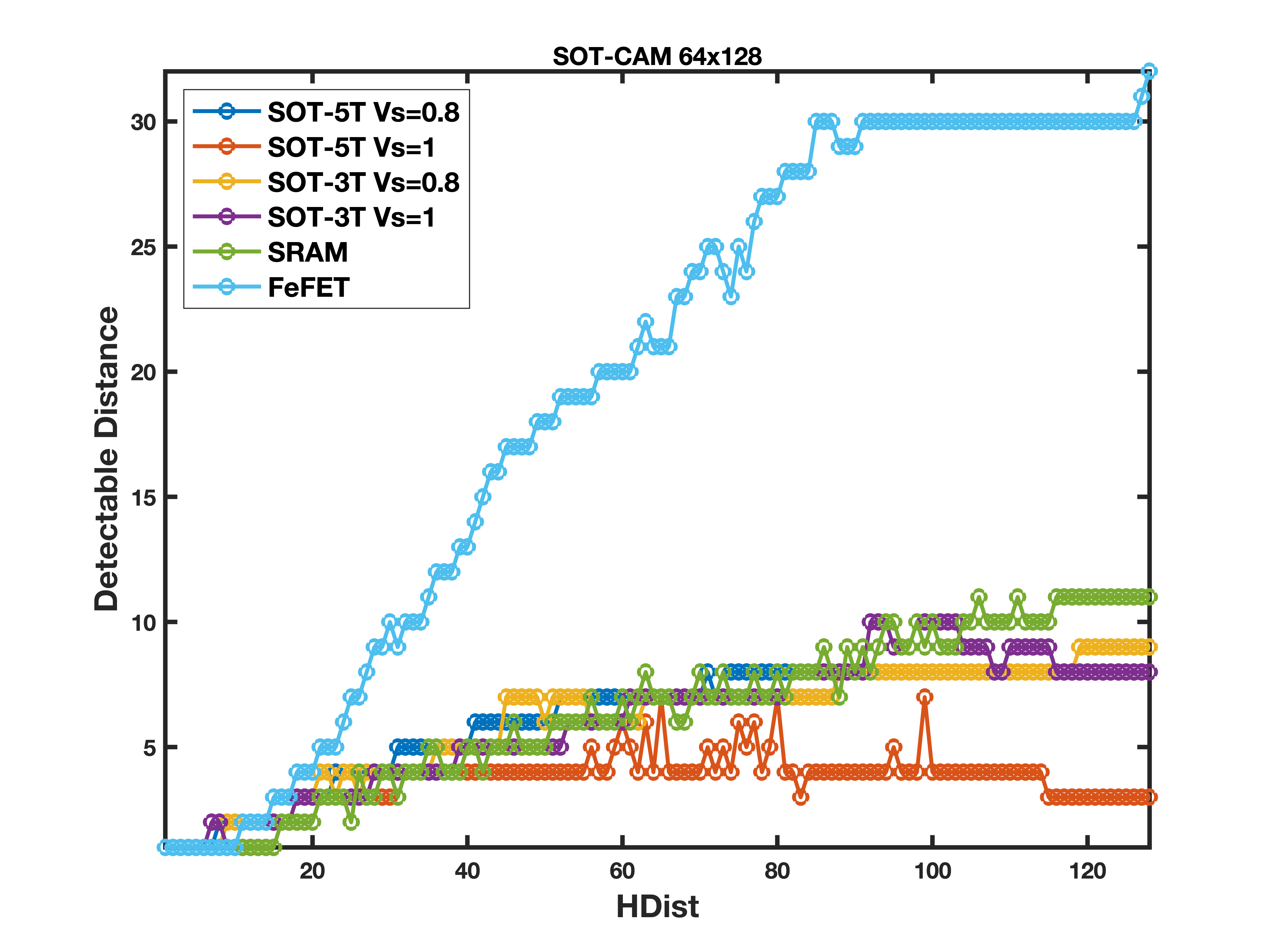}
	\caption{Minimum detectable distance for 5T and 3T SOT-CAM with Vs=0.8 V and 1 V, SRAM-CAM, and FeFET-CAM arrays.}
	\label{fig_mdd_bench}
\end{figure}

\section{Application-level Evaluation and Benchmarking}
To evaluate and benchmark the application-level accuracy and performance, we compare the proposed 5T SOT-CAM design with the previously proposed 3T SOT-CAM design \cite{narla2022design} along with the FeFET \cite{yin2018ultra} and SRAM based CAM \cite{pagiamtzis2006content} designs using layout extracted netlists\cite{narla2024cross}.
In all the cases, complementary data is stored in a CAM cell, the matchlines are precharged to Vdd (0.7 V) before evaluation and the searchlines (SBL and SBLB) are driven to Vs/0 depending on the search data.
For the 7nm FeFET-CAM, we use an FeFET with memory window of 0.46 V which has been reported in \cite{choe2021variability}. The capacitance of ferroelectric layer is calculated considering the ferroelectric layer thickness of 5nm and a dielectric coefficient of 35 \cite{hsu2020theoretical}. 
\subsection{Array-level Results}
Fig. \ref{fig_mdd_bench} compares the minimum detectable distance for various CAM designs. The FeFET CAM resolution suffers due to the FeFET capacitance which contributes to significant RC delay. The SOT-3T case uses a much larger LRS resistance value of 1 M$\Omega$; hence, the design has a lower IR drop and a larger RC delay than the SOT-5T case. When Vs=0.8 V, the SOT-3T design has a better resolution than the SOT-5T design. As the HDist increases, \cite{narla2024cross} shows that the difference in the consecutive delays drops which worsens the resolution. With the increasing HDist values, the delay values for the SOT-5T case drop at a steeper rate in comparison to the SOT-3T case where due to the large RC delay for charging the gate of the discharge transistor, ML discharge starts at a lower $\mathrm{V_{sot}}$ value. This means that the difference in the discharge delays of the consecutive Hamming distances decrease at a slower rate for the SOT-3T case than for the SOT-5T case, which is why the former has better resolution than the latter.

When Vs is increased to 1V, the resolution for the SOT-5T case improves because the difference between the discharge delays for the rows furthest and closest to the driver decreases. However, for the SOT-3T case, the RC delay dominates the ML discharge delay especially at the larger Hamming distance values. This prevents the difference in delays between the rows closest and furthest to the driver from shrinking as much as it does for the SOT-5T case when the search voltage is increased. Thus, for larger search voltages, the SOT-3T case shows lower improvement in the resolution than the SOT-5T case. 
Table \ref{tab_nns_bench} shows the results for a fixed radius search with HDist=20. The precision is better for the designs with better resolution, i.e. lower detectable distance.
\subsection{Application-level Results}
To benchmark the results at the application-level in comparison to the other state-of-the-art designs, we look at a sequential recommendation system where the similarity search is implemented using CAM arrays.
We use the sequential recommendation model from \cite{sasrec}. In this model, a self-attention block is used to predict the embedding of the next item that should be recommended to a user. In \cite{sasrec}, they use dot product ranking (DPR) to find the top-k items that are closest to the predicted item embedding. Using CAMs for ranking can be very costly as it would require ML discharge delay ranking to rank the top-k items which needs a significant amount of peripherals. Instead, CAMs are used for candidate generation with fixed-radius near neighbor search and these candidates are passed to the ranking stage to rank the top-k items. In this way, one can significantly reduce the number of items that need to go through DPR which is computationally expensive. We use the MovieLens 1M dataset from \cite{movielens} to train and test our model. Item embeddings are stored over multiple 64x128 CAM arrays using LSH encoding \cite{ni2019ferroelectric} with 128 bits. The attention model was trained for 100 epochs. During inference we used a test case with 1000 randomly selected negative items and 1 ground truth next item using the strategy from \cite{koren2008factorization}. HR@10 counts the fraction of times that the ground-truth next item is among the top 10 items after ranking for all valid test users. Our results show that with the use of CAMs, the final HR@10 (Table \ref{tab_sasrec_bench}) achieved are the same as those using DPR for the entire test set (0.32) while requiring fewer dot product operations.

Table \ref{tab_sasrec_bench} shows the results of using SOT-5T, SOT-3T, SRAM, and FeFET based CAMs for fixed-radius near neighbor search for candidate generation. The improvement in the number of dot product ranking operations is inversely proportional to the pool size of the candidates. The number of candidates outputted by the various CAM designs follows a pattern similar to the trend in resolution. FeFET-CAM shows the least amount of improvement. Larger search voltages work better for the SOT-3T case and more so for the SOT-5T case.
\subsection{Energy and Delay Results}
\par
Tables \ref{tab_edp_exact} and \ref{tab_edp_approx} show the search energy and delay for various designs for exact search and approximate search, respectively. The search delay for the SOT-3T case is higher than that for the SOT-5T case for larger Hamming distances since the RC delay in the circuit keeps the delay high as compared to the SOT-5T case where delay reduction with increasing Hamming distance is steeper. Increasing search voltages reduces delay due to larger $\mathrm{V_{sot}}$ mismatch voltage and as seen previously, improves the quality of search. Due to the large reduction in the delay, the search energy for Vs=1V decreases despite the increase in the search voltage. The energy consumption for the SOT-5T case is larger than the SOT-3T case due to the lower LRS resistance value used in the former design. The SOT based designs have larger search energy values compared to the SRAM and FeFET based designs since the former have large currents flowing through their arrays while the SRAM and FeFET based designs only require charging the gates of the transistors or FeFETs.

%% file: tex/conclusion.tex
\section{Conclusion}
In summary, we have presented a novel non-volatile spin transfer torque assisted spin-orbit torque based ternary content addressable memory with 5 transistors and 2 magnetic tunnel junctions. By using an STT-assisted write process, the design eliminates the need for a magnetic field for the write operation, therefore, improving magnetic immunity in comparison to the previous SOT-CAM design which used magnetic field-assisted write. We have performed a comprehensive study of the proposed design in terms of write, exact search, and approximate search. To accurately account for the impact of various sources of circuit parasitics at  advanced nodes like 7nm, we have used SPICE circuit simulations with layout extracted parasitics for bit-cells. We optimized our layout, array size, and search voltages to ensure accurate search operations. We project a search error rate for exact search operations lower than  3.9x10\textsuperscript{-9}\%, when various sources of variation are considered. For the approximate search, we show that the SOT-5T case with a search voltage of 1V have the best resolution amongst its counterparts.  Our results show that moving to STT-assisted SOT write operation to improve the magnetic immunity comes at the cost of a 1.3x increase in the area and 7x increase in the search energy. Finally, we benchmarked our design against SOT-3T, SRAM, and FeFET-based CAM designs using a CAM-based recommendation system, where our design shows 4.88x improvement in the operational speedup.

%% file: tex/ack.tex
\section*{Acknowledgment}
The authors gratefully thank Professors D. Ralph, S. X. Wang, and Drs. S. Dutta, and V. Kumar for insightful discussions.